\documentclass[pra,reprint,showpacs,aps,floatfix,amsmath,amssymb,amsfonts]{revtex4-1}

\usepackage{amsmath}
\usepackage{graphicx}
\usepackage{color}
\usepackage{inputenc}

\graphicspath{{figures/}}

\newcommand{\beq}{\begin{equation}}
\newcommand{\eeq}{\end{equation}}

\newcommand{\sumqp}{\sum_{\mathbf{q}\neq 0}}
\newcommand{\sumRp}{\sum_{\mathbf{R}\neq 0}}
\newcommand{\sumR}{\sum_{\mathbf{R}}}
\newcommand{\sumq}{\sum_{\mathbf{q}}}
\newcommand{\bk}{\mathbf{k}}
\newcommand{\bq}{\mathbf{q}}
\newcommand{\bp}{\mathbf{p}}

\newcommand{\eref}[1]{(\ref{#1})}

\begin{document}

\title{Beyond-mean-field corrections for dipolar bosons in an optical lattice}
\author{Jan Kumlin$^1$}
\author{Krzysztof Jachymski$^1$}
\author{Hans Peter B\"{u}chler$^1$}
\affiliation{
$^1$ Institute for Theoretical Physics III \& Center for Integrated Quantum Science and Technology (IQST), University of Stuttgart, Pfaffenwaldring 57, 70550 Stuttgart, Germany
}
\date{\today}

\begin{abstract}
Recent experiments with ultracold lanthanide atoms which are characterized by a large magnetic moment have revealed the crucial importance of beyond-mean-field corrections in understanding the dynamics of the gas. We study how the presence of an external optical lattice modifies the structure of the corrections. We find that deep in the superfluid regime the equation of state is well described by introducing an anisotropic effective mass. However, for a deep lattice we find terms with anomalous density dependence which do not arise in free space. For a one-dimensional lattice, the relative orientation of the dipole axis with respect to the lattice plays a crucial role and the beyond-mean-field corrections can be either enhanced or suppressed.
\end{abstract}

\maketitle

\section{Introduction}
Dilute gases of weakly interacting ultracold bosons are commonly described in terms of mean-field theory which predicts the formation of a Bose-Einstein condensate (BEC) described by a macroscopic wave function. The presence of long-range interactions can strongly affect the properties of the gas. Dipolar interactions are both long range and anisotropic, which brings in a number of interesting effects and possible applications~\cite{Lahaye:2009,baranov2012condensed}. For example, the partly attractive nature of the interaction can lead the gas to collapse~\cite{Lahaye2008}, which can be experimentally controlled by tuning the strength of the short-range repulsion by means of Feshbach  resonances~\cite{ChinRMP}. A confined strongly dipolar gas can exhibit a roton-maxon excitation spectrum~\cite{Santos2003,Ronen2007}, which has been experimentally observed~\cite{Chomaz2018}.

In recent years there has been a tremendous progress in experimental investigations of strongly dipolar systems, enabled by the realization of BEC of lanthanide atoms: erbium and dysprosium~\cite{Lu2011,Aikawa:2012}. It has then been discovered that when the dipolar gas is close to the stability boundary, it can form a set of stable dense droplets instead of collapsing~\cite{Kadau2016}. This effect can be explained by effective many-body repulsion induced by the leading beyond-mean-field corrections. These corrections are known as Lee-Huang-Yang (LHY) corrections in the case of contact interactions~\cite{LHY1957,Beliaev1958, Hugenholtz1959} and have also been extended to dipolar interactions~\cite{Lima2012}. The physics of these dipolar droplets has been the subject of intense studies~\cite{Kadau2016, Barbut2016,Wachtler2016,Baillie2016,Schmitt2016,Chomaz2018}. In the case of a single-component contact interacting Bose gas, the beyond-mean-field corrections are rather weak, however, those terms become dominant when the contribution from the mean-field term vanishes as in the case of partially attractive two-species mixtures~\cite{Petrov2015, Cabrera2018} and dipolar gases~\cite{Barbut2016}. Recently, the behavior of the beyond-mean-field corrections in the case of a dimensional crossover from three to low dimensions has been studied for contact~\cite{Ilg2018,Zin2018} and dipolar~\cite{Edler2017} interactions.

Optical lattices are an extremely useful tool for manipulation and control of ultracold gases~\cite{Jaksch1998,Lewenstein2007,Bloch2008}. They allow for creating a perfectly periodic external potential for the atoms with tunable depth and geometry. This enabled the experimental realization of the Bose-Hubbard model and demonstration of the quantum phase transition between the superfluid and Mott insulator phase~\cite{Greiner2002}. The interplay of dipolar interactions and optical lattice confinement gives rise to a variety of phenomena induced by the long-range nature of the interactions as well as their anisotropy~\cite{Goral2002,Buchler2007,Capogrosso2010,Sowinski2012,Wall2013}.

In this paper, we study the influence of an optical lattice on the beyond-mean-field corrections for a dipolar Bose gas. In particular, we investigate whether and how one can control the strength of the quantum fluctuations in a suitable way. We calculate the LHY correction for the case of a deep three-dimensional lattice as well as weak one-dimensional lattice. We find that in general the presence of the lattice enhances the fluctuations, but also introduces a nontrivial density dependence. Moreover, manipulating the relative orientation of the lattice and the dipole axis allows for controlling the strength of the LHY term.

This paper is structured as follows. In Section~\ref{sec:deep}, we consider the case of a deep three-dimensional cubic lattice. We compute the LHY correction to the ground state energy of an interacting Bose gas for contact (on-site) interactions, as well as for dipolar density-density interactions neglecting the interaction-induced tunneling effects. We show the emergence of an effective mass in the limit of a large healing length and calculate the corrections arising for larger interaction strength. In Section~\ref{sec:weak}, we turn to the case of weak one-dimensional lattice. We show that by manipulating the orientation of the lattice with respect to the dipole orientation axis it is possible to tune the magnitude of the LHY correction.

\section{Deep optical lattice}
\label{sec:deep}

\begin{figure}[h]
\includegraphics[width=0.25\textwidth]{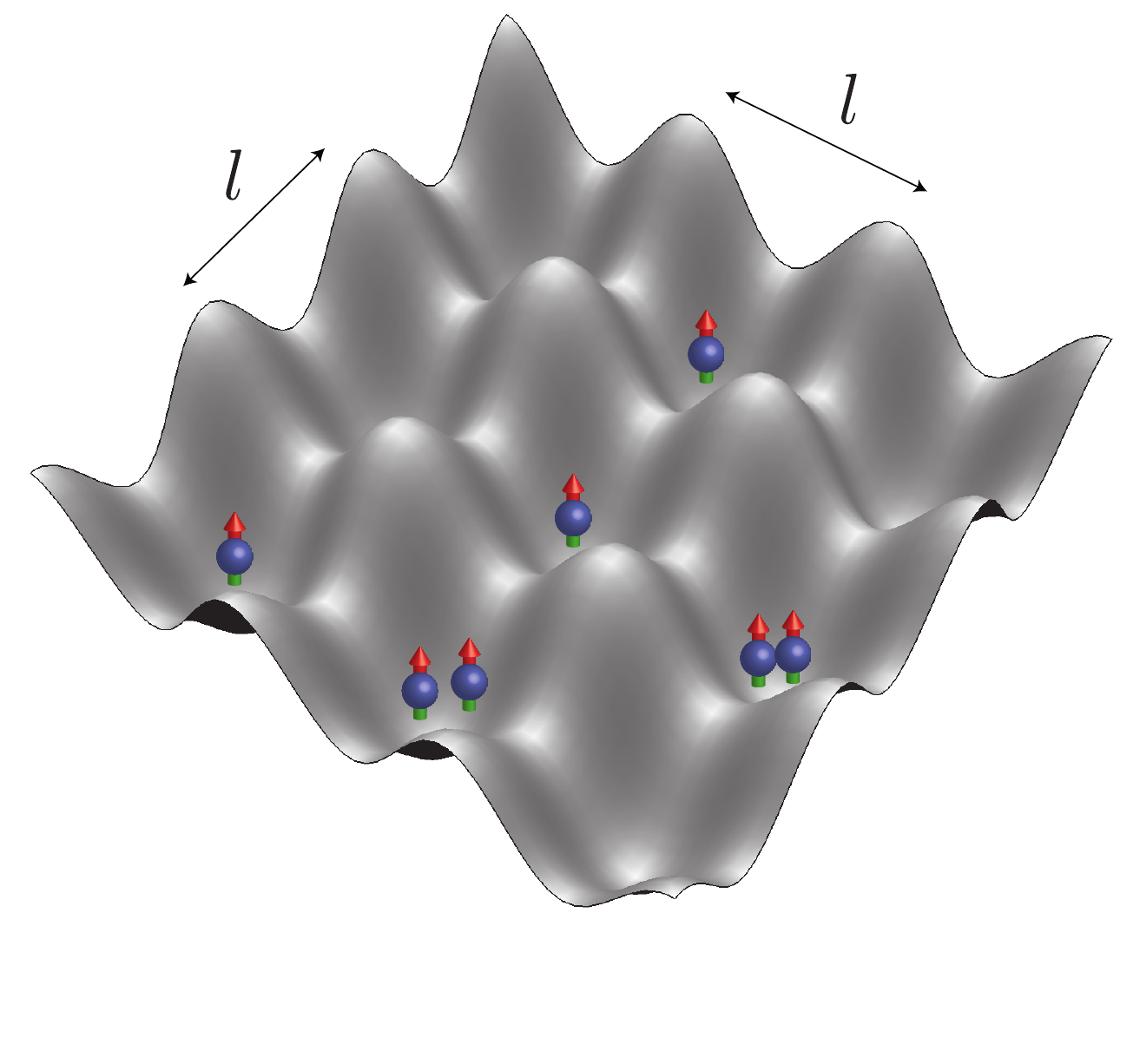}
\caption{\label{figdeep} Schematic illustration of the first studied case. Particles are trapped in a deep optical lattice in each direction, but are weakly interacting and remain superfluid.}
\end{figure}

\subsection{Hamiltonian}

We start our considerations with the general many-body Hamiltonian of an interacting Bose gas at zero temperature in second quantized form,
\begin{align}
H &= \int d^3 r \, \Psi^\dagger(\mathbf{r}) \left( - \frac{\hbar^2 \nabla^2}{2m} + U(\mathbf{r}) \right) \Psi(\mathbf{r}) \nonumber \\
& \qquad  + \frac{1}{2} \int d^3 r \int d^3 r' \, \Psi^\dagger(\mathbf{r}) \Psi^\dagger(\mathbf{r}') V(\mathbf{r} - \mathbf{r}') \Psi(\mathbf{r}') \Psi(\mathbf{r})\, . 
\label{eq:hamiltonian1}
\end{align}
In this Hamiltonian, $U(\mathbf{r})$ describes an external potential, e.g. a trapping potential or an optical lattice, while $V(\mathbf{r})$ describes the interaction potential (see Fig.~\ref{figdeep}). For dipolar particles which are polarized along one direction, the interaction potential can be represented in terms of a pseudopotential~\cite{Yi2001}
\begin{equation}
V(\mathbf{r}) = g_0 \delta(\mathbf{r}) +  \frac{C_{dd}}{4 \pi} \frac{1 - 3 \cos^2 \theta}{r^3}\, , 
\end{equation}
where $\theta$ is the angle between the direction of the polarization of the dipoles and their relative position. The first term of this pseudopotential accounts for the $s$-wave interactions which result from the short-ranged van-der-Waals interactions but contains also the contribution of the dipolar part of the potential. The second part stands for the higher partial waves, which are dominated by the long-range and anisotropic dipolar interaction. The coupling constant $g_0$ is related to the $s$-wave scattering length $a_s$ via  $g_0 = 4 \pi \hbar^2 a_s / m$.  The dipolar coupling constant $C_{dd}$ is $\mu_0 \mu^2 $ for particles having a magnetic dipole moment with $\mu_0$ being the permeability of the vacuum, and $d^2/ \epsilon_0$ for particles having an electric dipole moment $d$ with $\epsilon_0$ being the permittivity of the vacuum. In free space, the relative strength of the contact and the dipolar interaction is determined by the dimensionless parameter $\varepsilon_{dd}^0 = a_{dd} / a_s$ with the so-called dipolar length $a_{dd} = m C_{dd} / 12 \pi \hbar^2$.

Now, we consider the particles to be trapped in a deep three-dimensional simple cubic optical lattice described by the trapping potential $U(\mathbf{r}) = U_L \sum_{i = x, y, z} \sin^2(q_L r_i)$, where $U_L$ is the lattice depth and $ q_L = 2 \pi/ \lambda$ is the lattice vector with $\lambda$ being the laser wavelength and the lattice period is given by $l  = \lambda/2$. The typical energy scale of a particle in a lattice is the recoil energy $E_R = \hbar^2 q_L^2 / 2m$ and the strength of the lattice can be characterized by the dimensionless parameter $s = U_L/ E_R$. In the case of a deep lattice, we have $s \gg 1$. In this regime, one can assume that only the lowest Bloch band is occupied (in practice this is a good approximation already for $s \sim 10$~\cite{Bloch2008}) and we can rewrite the field operators $\Psi(\mathbf{r})$ and $\Psi^\dagger(\mathbf{r})$ in terms of the Wannier functions $w(\mathbf{r} - \mathbf{r}_i)$ localized around the lattice site $i$ at position $\mathbf{r}_i$:
\begin{equation}
\Psi(\mathbf{r}) = \sum_i w(\mathbf{r} - \mathbf{r}_i) a_i
\end{equation}
with the bosonic annihilation operator at lattice site $i$, $a_i$. The annihilation operators in real and quasi-momentum space are connected via a discrete Fourier transform
\begin{equation}
a_i = \frac{1}{\sqrt{N_L}} \sum_{\bk \in K} a_\bk e^{- i \mathbf{k}\mathbf{r}_i} \, ,
\end{equation}
where $N_L$ denotes the number of lattice sites and the summation over $\mathbf{k}$ is restricted to the first Brillouin zone $K$. Using these transformations, the single-particle term of the Hamiltonian becomes diagonal and the total Hamiltonian can be written as
\begin{equation}
H = \sum_\mathbf{k} \epsilon_\bk a_\bk^\dagger a_\bk + \frac{1}{2}\sum_{i,j,l,m} V_{ijlm} a^\dagger_i a^\dagger_j a_l a_m \, . 
\label{eq:hamiltonian_deep}
\end{equation}

The dispersion relation of a particle in a deep lattice is given by
\begin{equation}
\epsilon_\bk = -2t \left( \sum_{i = x,y,z} \cos(k_i l) - 3 \right)\, ,
\label{eq:single_particle_dispersion_deep}
\end{equation}
where the hopping amplitude $t$ is related to the lattice depth $U_L$, the recoil energy $E_R$, and the lattice spacing $l$ \cite{Jaksch1998}. Note that the zero of energy is shifted such that $\epsilon_0 = 0$. In the interaction part of \eref{eq:hamiltonian_deep}, the matrix elements $V_{ijlm}$ are given in terms of the Wannier functions
\begin{align}
V_{ijlm} &= \int d^3 r\, \int d^3 r' \, w^*(\mathbf{r} - \mathbf{r}_i) w^*(\mathbf{r}' - \mathbf{r}_j) V(\mathbf{r} - \mathbf{r}')  \nonumber \\
& \qquad \times w(\mathbf{r}' - \mathbf{r}_l) w(\mathbf{r} - \mathbf{r}_m) \, .  
\label{eq:matrix_elements}
\end{align}

For deep lattices, the Wannier functions are well localized and the contribution due to the contact interaction is only significant if $i = j = l = m$, such that we may write
\begin{equation}
V^\text{contact}_{ij} =  \delta_{ij} \, g_0 \int d^3 r \, \vert w (\mathbf{r}) \vert^4 \equiv g \delta_{ij}\, .
\end{equation}
Approximating the Wannier function at a given lattice site by the ground state wavefunction of an harmonic oscillator, $g$ can be calculated explicitly as~\cite{Jaksch1998}
\begin{equation}
g  = g_0 \frac{(2\pi)^{3/2}}{l^3} s^{3/4} \, . 
\label{eq:g_harmonic}
\end{equation}

For the dipolar part, we replace the Wannier functions by $\delta$ functions,
\begin{equation}
w^*(\mathbf{r} - \mathbf{r}_i) w(\mathbf{r} - \mathbf{r}_m) \approx \delta_{im} \delta (\mathbf{r} - \mathbf{r}_i)\, . 
\end{equation}
The matrix elements now only depend on sites $i$ and $j$ and are given by
\begin{equation}
V_{ij} = g \delta_{ij} + \frac{C_{dd}}{4\pi} \frac{1 - 3 \cos^2 \theta_{ij}}{ \vert \mathbf{r}_i - \mathbf{r}_j \vert^3} \, , 
\end{equation}
with $\theta_{ij}$ being the angle between sites $i$ and $j$. Note that the on-site contribution from the dipolar part is zero for an isotropic confinement at each lattice site. The Hamiltonian (\ref{eq:hamiltonian_deep}) then reduces to
\begin{equation}
H = \sum_\mathbf{k} \epsilon_\bk a_\bk^\dagger a_\bk + \frac{1}{2}\sum_{i,j} V_{ij} n_i n_j \, . 
\end{equation}
Taking into account the spatial structure of the Wannier states and computing the matrix elements in \eref{eq:matrix_elements} explicitly gives rise to additional processes such as density-assisted and correlated tunneling~\cite{Sowinski2012, Wall2013}. These processes are strongly suppressed for deep lattices due to the exponential decay of the Wannier functions, but can lead to additional corrections for moderate lattice depths. The role of these terms is discussed in Appendix~\ref{app:nextorder}.

For what follows, we also need to transform the interaction part of (\ref{eq:hamiltonian_deep}) into momentum space which requires the discrete Fourier transform of the interaction potential and might be written as
\begin{equation}
V(\mathbf{k}) \equiv V_\mathbf{k} = g \left( 1 + \varepsilon_{dd} \frac{3}{4\pi} \sum_j e^{i \mathbf{k} \mathbf{r}_j} \frac{1 - 3 \cos^2 \theta_j}{\vert \mathbf{j} \vert^3} \right) \, ,
\label{eq:ft_potential}
\end{equation}
where the parameter $\varepsilon_{dd}$ is renormalized by the lattice and related to its free space variant by
\begin{equation}
\varepsilon_{dd} = \frac{ \varepsilon_{dd}^0}{l^3 \int d^3 r \, \vert w(\mathbf{r} ) \vert^4 } = \frac{ \varepsilon_{dd}^0}{(2\pi)^{3/2} s^{3/4}}\, . 
\end{equation}
The last result was obtained using Eq.~(\ref{eq:g_harmonic}). The exact form of $V_\mathbf{k}$ can be obtained analytically under our approximations (see Appendix~\ref{app:appendix_dipolarFT}) and leads to noticeable differences with respect to the free-space Fourier transform in which $V_\mathbf{k}$ only depends on the angle between $\mathbf{k}$ and the direction of the polarization of the dipoles but not on the magnitude of $\mathbf{k}$.

\subsection{Beyond-mean-field corrections}

In order to calculate the beyond-mean-field energy corrections, we restrict ourselves to the case where the system is in the superfluid phase and the mean-field approach correctly describes its properties. The correction to the mean-field energy is then given by~\cite{Hugenholtz1959}
\begin{equation}
\label{eq:pines}
\frac{E^{(2)}}{V} - \frac{1}{2} n \mu^{(2)} = \frac{1}{2} \int_K \frac{d^3 k }{(2\pi)^3} \frac{(\epsilon_\bk + n V_\bk - E_\bk) (\epsilon_\bk - E_\bk)}{2E_\bk}\, ,
\end{equation}
where $V = N_L l^3$ is the volume of the system, $n = N/V$ the density, and $\mu^{(2)}$ denotes the second-order correction to the chemical potential, which is related to the energy by $\mu = d E / d n$. The Bogoliubov dispersion relation $E_k$ is given by $E_\bk = \sqrt{\epsilon_\bk (\epsilon_\bk + 2 n V_\bk)}$ with the non-interacting single-particle dispersion $\epsilon_\bk$ [\ref{eq:single_particle_dispersion_deep}] and the (discrete) Fourier transform of the dipolar potential, $V_\bk$. We note that as the integration is restricted to the first Brillouin zone, the result is in principle always convergent and no renormalization is needed. We will now study the structure of the correction for different cases, starting for simplicity with the contact interactions.

\subsubsection{Contact interaction}
For a purely contact interacting Bose gas, we have $V_\bk = g$ and thus no momentum dependence of the interaction potential. In order to simplify the integral on the right-hand side of \eref{eq:pines} and to make it dimensionless, we introduce the effective mass $m^* = \hbar^2 / (2 t l^2)$ and the healing length $\xi^2 = \hbar^2 / (2 m^* n g)$. We further introduce the dimensionless parameter $\alpha = \xi^2 / l^2 = t/ng$, which parametrizes the relative strength of the interaction. Staying in the superfluid phase requires $ \alpha \gg 1$. Using the substitution $k_i l = u_i l / \xi = u_i / \sqrt{\alpha}$, the integral in \eref{eq:pines} reduces to
\begin{equation}
\frac{1}{2} n g \frac{1}{l^3 \alpha^{3/2}} \int_{-\pi \sqrt{\alpha}}^{\pi \sqrt{\alpha}} \frac{d^3 u}{(2 \pi)^3} \frac{(\epsilon_\mathbf{u} + 1 - E_\mathbf{u})(\epsilon_\mathbf{u} - E_\mathbf{u})}{2 E_\mathbf{u}}
\end{equation}
with $\epsilon_\mathbf{u} = -2 \alpha \left( \sum_{i=x,y,z} \cos( u_i / \sqrt{\alpha}) - 3 \right)$ and $E_\mathbf{u} = \sqrt{\epsilon_\mathbf{u} ( \epsilon_\mathbf{u} + 2)}$. The prefactor $ng / (\alpha^{3/2} l^3)$ in front of the integral can equivalently be written as
\begin{equation}
\frac{ng}{l^3 \alpha^{3/2}} = \frac{(ng)^{5/2} (2m^*)^{3/2}}{\hbar^3}\, .
\end{equation}
This is, up to a constant numerical factor, exactly the form of the LHY correction in the absence of the optical lattice such that its effects are contained solely in the integral
\begin{equation}
I(\alpha) = \frac{1}{2} \int_{-\pi / \sqrt{\alpha}}^{\pi / \sqrt{\alpha}} \frac{d^3 u}{(2\pi )^3} \frac{(\epsilon_\mathbf{u} + 1 - E_\mathbf{u})(\epsilon_\mathbf{u} - E_\mathbf{u})}{2 E_\mathbf{u}}\, . 
\end{equation}
In the limit $\alpha \to \infty$, which corresponds to free space, the integral $I(\alpha)$ can be calculated analytically and approaches the constant $I(\alpha \to \infty) = - 1 / 15 \sqrt{2} \pi^2$ such that
\begin{equation}
\frac{E^{(2)}}{V} - \frac{1}{2} n \mu^{(2)}  = \frac{(ng)^{5/2} (2m^*)^{3/2}}{\hbar^3} \left( - \frac{1}{15 \sqrt{2} \pi^2} \right) \, . 
\end{equation}
Solving the differential equation with the initial condition $E^{(2)}(n = 0) = 0$ ensuring the correct mean-field result, one recovers the well-known free-space LHY term
\begin{equation}
\frac{E^{(2)}}{V} = \frac{8}{15 \pi^2} \frac{(m^*)^{3/2} (ng)^{5/2}}{\hbar^3}\, 
\end{equation}
with an effective mass accounting for the underlying lattice structure. For finite $\alpha$, one can compute the integral numerically and the result is shown in Fig.~\ref{fig:I_dip_full}. As expected, there are corrections to the free-space value at finite $\alpha$ which introduces an additional density dependence to the right-hand side of \eref{eq:pines}. We discuss this additional density dependence and its influence on the beyond-mean-field energy corrections in Sec.~\ref{sec:structure_energy_correction}.

\subsubsection{Dipolar Interaction}

\begin{figure}
\centering
\includegraphics[width = 0.48\textwidth]{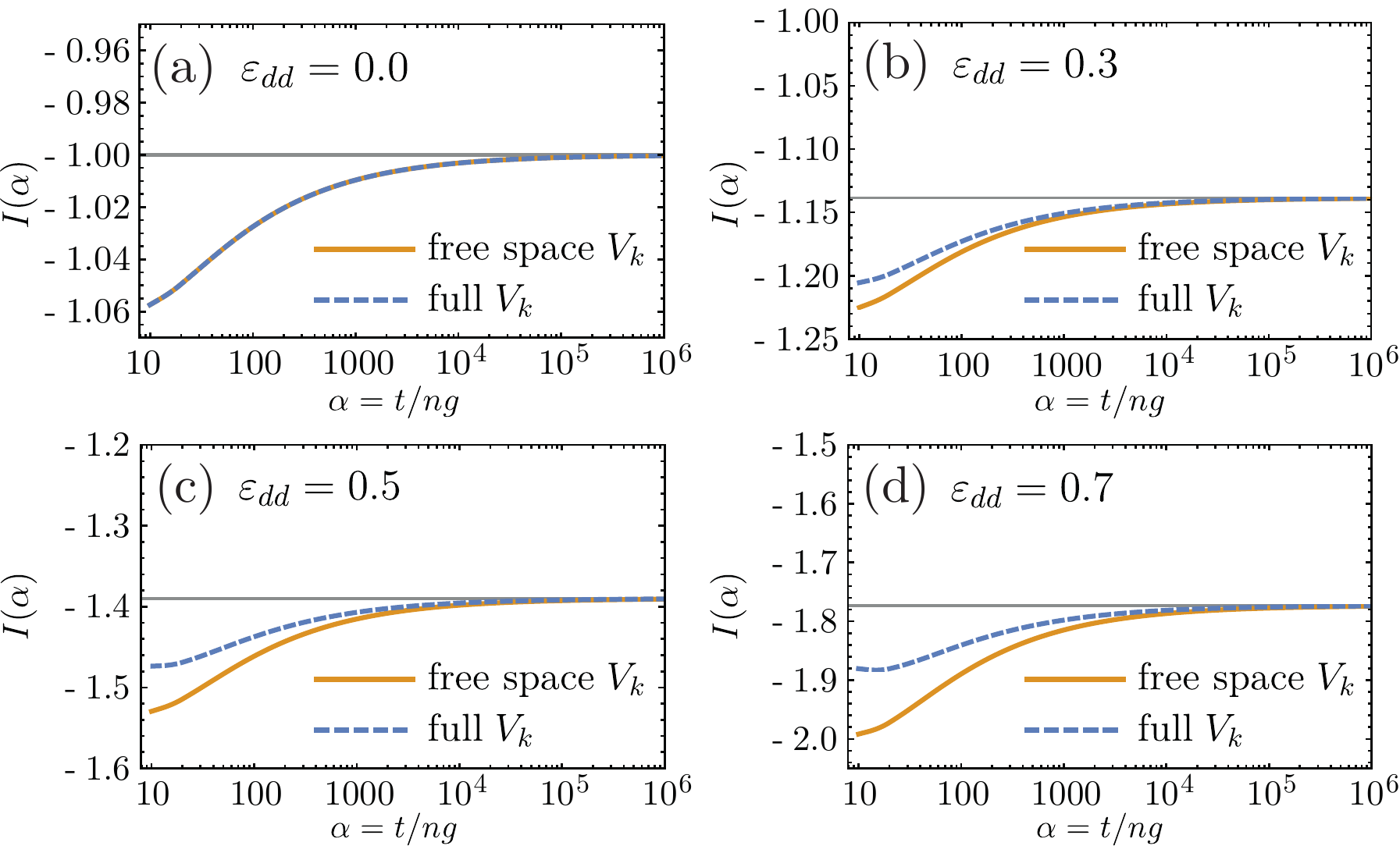}
\caption{$I(\alpha)$ coefficient for full [orange (solid) line] lattice dipolar interaction compared with the free-space one [blue (dashed) line] for several different values of (a) $\varepsilon_{dd} = 0.0$, (b) $\varepsilon_{dd} = 0.3$, (c) $\varepsilon_{dd} = 0.5$, and (d) $\varepsilon_{dd} = 0.7$. For the contact interaction (a), there is no difference between both cases as the interaction potential is constant in momentum space.}
\label{fig:I_dip_full}
\end{figure}

We now turn to the case of dipolar interactions. It is interesting to study both the differences which arise with respect to the contact interaction case as well as the effect of the lattice on the dipolar gas. In order to achieve better understanding of the role of the lattice, we first approximate the lattice Fourier transform of the dipolar interaction given in~\eqref{eq:vk_final} by the free-space Fourier transform
\begin{equation}
V_{dd}^{\text{free}} ( \mathbf{k}) = g \varepsilon_{dd} \left( 3 \frac{k_z^2}{k_x^2 + k_y^2 + k_z^2} - 1 \right)\, ,
\label{eq:free_space_FT}
\end{equation}
which only depends on the angle between $\mathbf{k}$ and the direction of polarization of the dipoles (which is assumed to be the $z$ axis). The integral $I(\alpha)$ can only be calculated numerically for finite $\alpha$ and the results for different values of $\varepsilon_{dd}$ are shown in Fig.~\ref{fig:I_dip_full}. The asymptotic value for $\alpha \to \infty$ can again be calculated analytically and reads as
\begin{equation}
I(\alpha \to \infty) = - \frac{1}{15 \sqrt{2} \pi^2} F(\varepsilon_{dd}) \,
\end{equation}
with $F(\varepsilon_{dd}) = \frac{1}{2} \int_{-1}^1 du \, (1 + \varepsilon_{dd} (3u^2 - 1))^{5/2}$ accounting for the anisotropic nature of the dipolar interaction (see also~\cite{Lima2012}). In the absence of dipolar interactions, it reduces to $F(0) = 1$. In the case of finite $\alpha$, the integral $I(\alpha)$ leads to corrections qualitatively similar as in the case of contact interaction, while their magnitude increases with increasing $\varepsilon_{dd}$. 

\begin{figure}
\centering
\includegraphics[width=0.45\textwidth]{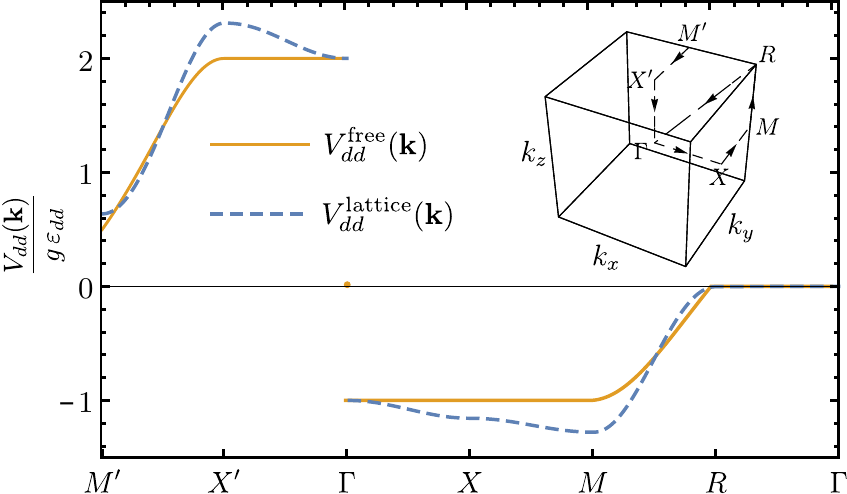}
\caption{Comparison between free space [orange (solid) line] and lattice [blue (dashed) line] dipolar Fourier transform plotted over the first Brillouin zone.}
\label{fig:Vk_BZ}
\end{figure}

The effect of the dipolar interaction is even more enhanced when taking the full lattice Fourier transform of the dipolar interaction as derived in Sec.~\ref{app:appendix_dipolarFT}. For $\alpha \to \infty$, $I(\alpha)$ also approaches the free-space result and is equivalent to taking the free-space Fourier transform \eref{eq:free_space_FT}, whereas the deviations with respect to the case of contact interaction are more prominent for finite $\alpha$. This can be understood by comparing the lattice Fourier transform and the free-space Fourier transform as shown in Fig.~\ref{fig:Vk_BZ}. In contrast to free space, the lattice Fourier transform of the dipolar potential also depends on the absolute value of the momentum. For large values of $\alpha$, the integral in \eref{eq:pines} only gives a contribution near $\mathbf{k} = 0$ where there is little difference between both cases, while for decreasing values of $\alpha$, the integral also probes higher momenta where the lattice enhances the effect of the dipolar potential.

\subsubsection{General structure of the energy correction}
\label{sec:structure_energy_correction}

In the previous sections, we have seen that in the presence of a deep lattice, corrections to the usual free-space behavior arise. These corrections lead to a modified scaling in the density which is discussed in this section. 
In order to do so, we rewrite \eref{eq:pines} expressing the energy density $E_0^{(2)} / V $ in terms of $t/l^3$, that is  $E_0^{(2)}/V = (t / l^3) e_0^{(2)}$. Substituting $ n \to t / \alpha g$ and noting that $ d / dn = - (\alpha^2 t / g) \, d / d \alpha$, we end up with the dimensionless differential equation 
\begin{align}
e_0^{(2)} + \frac{1}{2} \alpha \frac{d e_0^{(2)}}{d \alpha} & = \alpha^{-5/2} I(\alpha) \nonumber \\
& = - \frac{F(\varepsilon_{dd})}{15 \sqrt{2} \pi^2} \alpha^{-5/2} \left[1 + f(\alpha)\right] \equiv P(\alpha)\, .
\label{eq:diff_eq}
\end{align} 
Note that in this expression, all contributions reminiscent of the free-space result are written explicitly and the function $f(\alpha)$ provides all the corrections coming from the lattice and $f(\alpha \to \infty) = 0$. The differential equation can formally be solved and using the initial condition $e_0^{(2)} (\infty) = 0$ such that we obtain the correct mean-field result, the solution reads as
\begin{equation}
e_0^{(2)} =  - \frac{1}{\alpha^{-2}} \int_\alpha^\infty dy\, 2 P(y) y\, . 
\label{eq:solution_diff_eq}
\end{equation}
This is the particular solution of \eref{eq:diff_eq}, while the homogeneous solution would only affect the mean-field energy in which we are not interested right now. From another point of view, one can also solve \eref{eq:pines} with an inhomogeneity $G(n)$ which leads to the solution
\begin{equation}
\frac{E_0^{(2)}}{V} = - n^2 \int_0^n dy \, \frac{2 G(y)}{y^3}\, . 
\end{equation}
Substituting $\alpha = t / ng$ and expressing the energy density in terms of $t / l^3$ leads to the solution \eref{eq:solution_diff_eq}. For the rest of the discussion, we return to the dimensionless form of the solution, \eqref{eq:solution_diff_eq}. In the presence of the lattice \footnote{The following discussion is essentially more general provided that there exists a dimensionless parameter such that the inhomogeneity has the same scaling as in free space in some limit of this parameter.}, we can split the inhomogeneous term on the right-hand side of the differential equation into one part containing the free-space result and one part containing the corrections arising from the lattice:
\begin{equation}
P(\alpha) = P_0(\alpha) + \Delta P(\alpha) = - \frac{F(\varepsilon_{dd})}{15 \sqrt{2} \pi^2} \alpha^{-5/2} \left[ 1 + f(\alpha)\right] \, . 
\end{equation}
The first term gives rise to the standard result in free space $ e_0^{(2),0} = \frac{4}{15 \sqrt{2} \pi^2} F(\varepsilon_{dd}) \alpha^{-5/2}$ which leads to the beyond-mean-field correction
\begin{equation} 
\frac{E_0^{(2)}}{V} = \frac{4 F(\varepsilon_{dd})}{15 \sqrt{2} \pi^2} \frac{(2m^*)^{3/2} (ng)^{5/2}}{\hbar^3}\, .
\end{equation}
For the second part, we see from Fig.~\ref{fig:I_dip_full} that we can describe the corrections due to the lattice by a function $f(\alpha) = c \alpha^{-\gamma}$ for $\alpha \gg 1$. The parameters $c$ and $\gamma$ will in general depend on the relative dipolar interaction strength $\varepsilon_{dd}$ and will later be determined by fitting the expression to the numerically calculated $I(\alpha)$. Note that $\gamma >0 $ as $f(\alpha \to \infty) = 0$. Thus, the energy correction due to the second part reads as
\begin{align}
\Delta e_0^{(2)} & = \frac{F(\varepsilon_{dd})}{15 \sqrt{2} \pi^2} \frac{2 c}{\alpha^3} \int_{\alpha}^{\infty} dy \, y^{-3/2 - \gamma} \nonumber \\
&= \frac{4 F(\varepsilon_{dd})}{15 \sqrt{2} \pi^2} \frac{c}{1+2\gamma} \alpha^{-5/2 - \gamma} \, .
\end{align}
Finally, the complete beyond-mean-field energy correction reads as
\begin{equation}
\frac{E_0^{(2)}}{V} = \frac{8}{15 \pi^2} \frac{(m^*)^{3/2} (ng)^{5/2}}{\hbar^3} F(\varepsilon_{dd}) \left( 1 + \frac{c}{1 + 2 \gamma} \left( \frac{ng}{t} \right)^\gamma \right)\, . 
 \label{eq:energy_correction_final}
\end{equation}
Since $\gamma > 0$, the correction to the beyond-mean-field correction to the ground state energy due to the lattice increases with increasing density. In the limit $t \gg ng$, \eref{eq:energy_correction_final} reduces to the free-space result with a renormalized mass as discussed before. For the present setup of a three-dimensional simple cubic lattice, we determine the coefficients $c$ and $\gamma$ by fitting to the results obtained using numerical integration. For the scaling coefficients, we get $\gamma \approx 1/2$ independent of $\varepsilon_{dd}$ and valid also for contact interactions, whereas the coefficient $c$ ranges from $c \approx 0.3 $ for $\varepsilon_{dd} = 0$ to $c \approx 0.76$ for $\varepsilon_{dd} = 0.7$. 




\section{Weak one-dimensional lattice}
\label{sec:weak}

Up to now, we have considered a three-dimensional optical lattice. However, an additional intriguing possibility is to restrict the lattice to one dimension and play with the relative orientation of the dipoles and the wave vector of the lattice. In this section, we demonstrate that this generates additional corrections to the usual beyond-mean-field corrections that can be enhanced or diminished depending on the relative orientation between the lattice and the dipoles. The basic assumption of our analysis is that the lattice is weak and can be treated as a perturbation to the free-space case and we do not restrict ourselves to the lowest Bloch band. 
\subsection{Model}

\begin{figure}
\includegraphics[width = 0.25\textwidth]{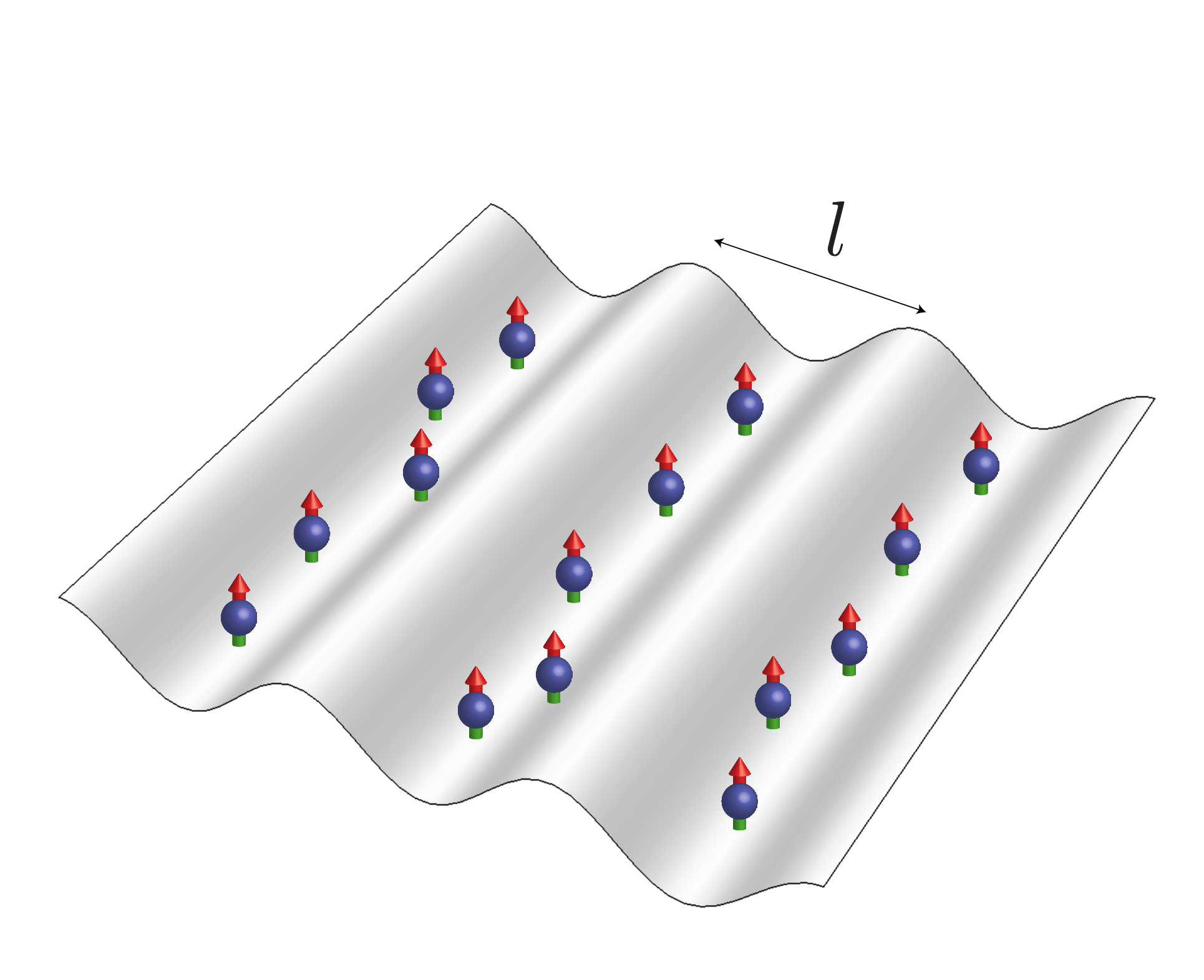}
\caption{\label{fig:weak} Schematic illustration of the second studied case. Particles are very weakly trapped by the lattice in one dimension only.}

\end{figure}

The one-dimensional lattice potential is given as
\begin{equation}
U(\mathbf{r}) = U_L \sin^2( \mathbf{q}_L \mathbf{r})\, ,
\end{equation}
where, as before, $q_L = \pi / l$ is the lattice vector and $l$ is the lattice period (see also Fig.~\ref{fig:weak}). The orientation of the lattice relative to the dipole axis, which we assume to be parallel to the $z$ axis, can be varied by varying the direction of the lattice vector $\mathbf{q}_L$. In the presence of this lattice, the eigenfunctions of the non-interacting problem are given in terms of a product of plane waves (in those directions where no lattice is present) and Bloch functions (in the direction of the lattice),
\begin{equation}
\psi_\bk(\mathbf{r}) = e^{i \mathbf{k}_\perp \mathbf{r}} \phi_{k_{\parallel}}(\mathbf{r}_{\parallel}) \, . 
\end{equation}
In this basis, the single-particle part of the Hamiltonian is diagonal, $H_0 = \sum_\bk \epsilon_\bk a_\bk^\dagger a_\bk$, where $\epsilon_\bk = \hbar^2 \mathbf{k}_\perp^2 / 2m + \tilde{\epsilon}_k$. The components of $\mathbf{k}$ that are parallel to the lattice are restricted to the first Brillouin zone, $k \in [-\pi / l, \pi/ l]$. In general, the lattice dispersion $\tilde{\epsilon}_k$ cannot be written down in closed form for arbitrary lattice depth while for weak lattices there are approximations (see Appendix~\ref{app:weak}). As for the deep lattice, we introduce the recoil energy $E_R = \hbar^2 q_L^2 / 2m$ and the dimensionless lattice depth $s = U_L / E_R$ which for a weak lattice is assumed to be much smaller than one, $s \ll 1$. 

\subsection{Contact interaction}
We proceed as in the previous section and start with the contact interaction where the interaction part of the Hamiltonian takes the form
\begin{equation}
H_{\rm int} = \frac{g}{2V} \sum_{\mathbf{k}, \mathbf{k}', \mathbf{q}} a^\dagger_{\bk+\mathbf{q}} a^\dagger_{\bk'-\mathbf{q}} a_\bk a_{\bk'} \, .
\end{equation}
We now apply the Bogoliubov theory following, e.g., Ref.~\cite{AGD}, which leads to the excitation spectrum
\begin{equation}
E_\bk = \sqrt{\epsilon_\bk (\epsilon_\bk + 2ng)}
\end{equation}
as well as the beyond-mean-field contributions to the ground-state energy per volume
\begin{equation}
\frac{\Delta E_0}{V} = \frac{1}{2V} \sum_\bk \left[E_\bk - (\epsilon_\bk + ng)\right] \, . 
\end{equation}
Since the lattice is assumed to be weak, the dispersion $\epsilon_\bk$ will only have small deviations from the dispersion in free space such that we can split the result into a term corresponding to the free space result, the well-known LHY term, and one additional term which includes all the corrections to it. The free space result diverges which can be cured by a proper renormalization of the coupling constant $g$ (see, e.g., Ref.~\cite{AGD}). We focus here on the remaining corrections to the LHY term due to the lattice which can be expressed as
\begin{equation}
\frac{\Delta E_0^s}{V} = \frac{1}{2} \int \frac{d^3 k }{(2\pi)^3} ( E_\bk^s - E_\bk^0 - (\epsilon_\bk^s - \epsilon_\bk^0))\, , 
\label{eq:corrections_weak_integral}
\end{equation}
where $E_\bk^s$ ($\epsilon_\bk^s$) and $E_\bk^0$ ($\epsilon_\bk^0$) denote the Bogoliubov excitation spectrum (single-particle dispersion) in the presence of the lattice and in free space, respectively.

Similarly to the case of a deep lattice, we introduce a dimensionless quantity
\begin{equation}
\beta^2 = \frac{E_R}{ng} = \pi^2  \frac{\xi^2}{l^2} = \pi^2 \alpha
\end{equation}
which compares the healing length $\xi$ with the optical lattice period $l$.

\begin{figure}
\centering
\includegraphics[scale=0.7]{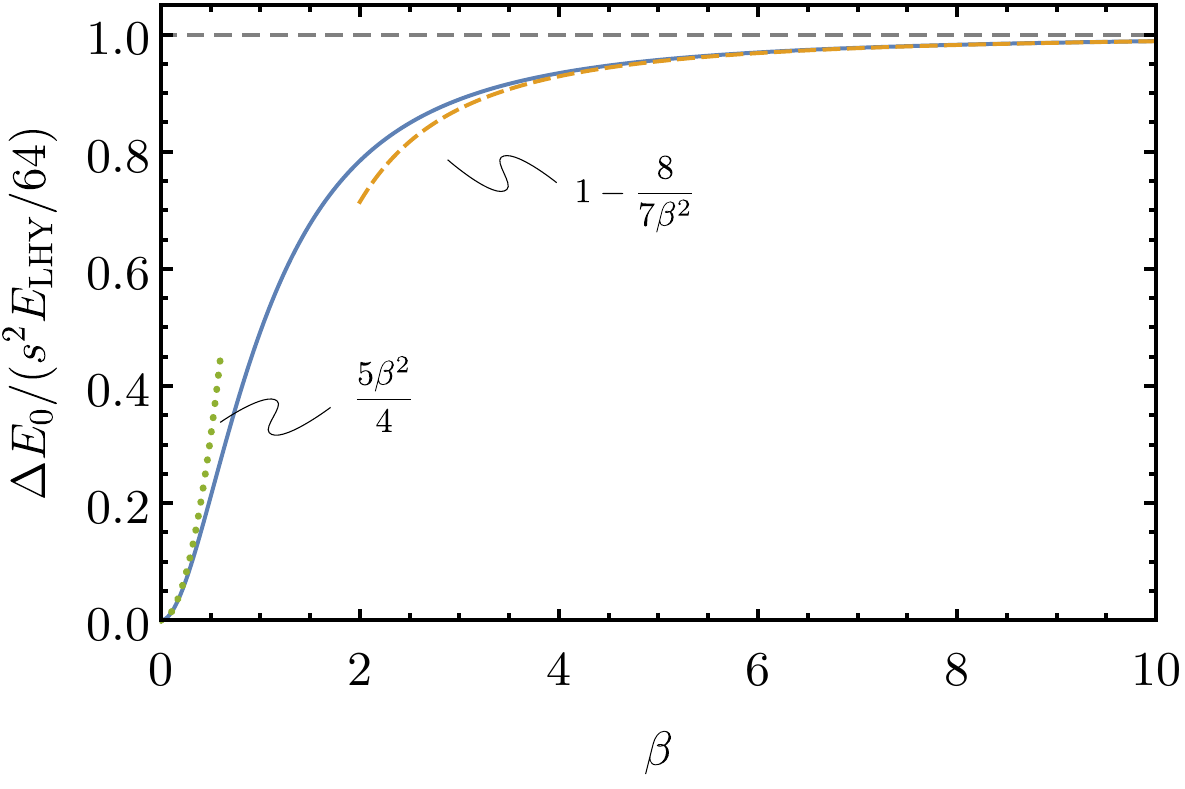}
\caption{\label{fig:correction_contact} [Blue (solid) line] General dependence of the beyond-mean-field correction to the energy on the parameter $\beta$. [Green (dotted) line] Asymptotic behavior for $\beta \to 0$. [Orange (dashed) line] Asymptotic behavior for $\beta \to \infty$.}
\end{figure}

As we are only interested in a weak lattice, we calculate the beyond-mean-field corrections due to the lattice (\ref{eq:corrections_weak_integral}) to leading order in $s$, for which analytical results can be obtained (see Appendix~\ref{app:weak} for further details). The general dependence of the beyond-mean-field corrections on the parameter $\beta$ in the case of a contact-interacting Bose gas is shown in Fig.~\ref{fig:correction_contact} [blue (solid) line] and also includes the leading behavior in the limits $\beta \ll 1$ [green (dash-dotted) line] and $\beta \gg 1$ [orange (dashed) line] which are discussed below.

In the limit of $\beta \to 0$, which corresponds to $l/\xi \gg 1$, the leading behavior of Eq.~(\ref{eq:corr_contact}) is given by
\begin{equation}
\frac{\Delta E_0^2}{V} = \frac{E_{\rm LHY}}{V} s^2 \beta^2 \frac{5}{256} \to 0 \, 
\end{equation}
with the usual LHY correction
\begin{equation}
\frac{E_{\rm LHY}}{V} = \frac{8}{15 \pi^2} \frac{m^{3/2} (ng)^{5/2}}{ \hbar^3} \, .
\end{equation}
The vanishing influence of the lattice in this case is intuitively clear as the limit $l / \xi \gg 1$ indeed should correspond to free space. 

In the opposite limit, $\beta \to \infty$ ($l / \xi \ll 1$), the correction instead approaches a constant and reads as
\begin{equation}
\frac{\Delta E_0^s}{V} = \frac{E_{\rm LHY}}{V} \frac{s^2}{64}\, . 
\end{equation}
Combining this with the mean-field energy, the ground state energy per volume is given by
\begin{equation}
\frac{E_0}{V} = \left( \frac{n^2 g}{2} + \frac{E_{\rm LHY}}{V} \right) \left( 1 + \frac{s^2}{64}\right) \, . 
\end{equation}
Both the mean-field term and the beyond-mean-field term are enhanced by the presence of the lattice in the same way in the leading order. The correction to the mean-field term comes from the correction to the $k = 0$ mode due to the coupling to modes with $k = \pm q_L$ in the presence of the lattice. 

Interestingly, this result can also be obtained when assuming that the bosons are in free space but acquire an anisotropic effective mass along the direction of the lattice (say in $z$ direction) with $m_z = m/(1 - s^2/ 32) = m/\gamma^2$ to leading order in $s$. This immediately leads to 
\begin{equation}
E_{\rm LHY} \to \gamma^{-1} E_{\rm LHY} \approx \left( 1 + \frac{s^2}{64}\right) E_{\rm LHY} \, . 
\end{equation}

Figure~\ref{fig:correction_contact} shows that the asymptotic value for $\beta \gg 1$ is already approached for $\beta$ on the order of 10. This suggests that describing the system as a free-space gas with an effective anisotropic mass should be adequate already at values of $\beta$ on the order of 10 which corresponds to $\xi \approx 3 l$. Typically, the lattice period is on the order of a few hundred nanometers while the healing length is on the order of micrometers for standard cold-atom experiments~\cite{Bloch2008} such that this approximation can be used in typical experiments.

\subsection{Dipolar interactions in the $\beta\to\infty$ limit}
After having discussed the case of a contact interacting gas, we turn to the the case where the particles interact via dipolar interactions given by the interaction potential
\begin{equation}
V_{dd} (\mathbf{r}) = g \epsilon_{dd} \frac{1 - 3 \cos^2 \theta}{r^3} \, ,
\end{equation}
where $\theta$ is the angle between the direction of the dipole moment, which is assumed to be parallel to the $z$ axis and the relative position of the two particles given by $\mathbf{r}$. As we are only concerned about the leading order corrections for small values of $s$, we can consistently neglect the weak lattice effect on the interaction potential and use the free-space formula 
\begin{equation}
V_\bk = g \left( 1 + \varepsilon_{dd} \left(3 \frac{k_z^2}{k_x^2 + k_y^2 + k_z^2} - 1 \right) \right) \equiv g \tilde{V}_\bk \, . 
\label{eq:ddi_ft}
\end{equation}
In contrast to the contact interacting case, the integrals cannot be solved analytically for arbitrary $\beta$. However, the previous analysis suggested to view the limit $\beta \to \infty$, or equivalently $ l / \xi \ll 1$, as nothing else but a gas in free space with an anisotropic effective mass. In the following, we make use of this simplification and derive the corrections due to the weak lattice potential along an arbitrary direction. In the end, we discuss two special limits when the lattice is parallel and perpendicular to the polarization axis of the dipoles. 

Since the dipole potential is invariant under rotation around the $z$-axis, the final results can only depend on the angle $\eta$ between the direction of the polarization of the dipoles, which is assumed to be the $z$-axis, and the lattice vector $\mathbf{q}_L$. For simplicity, we choose the lattice wave vector to be in the $yz$-plane. Introducing the effective mass $m_{\rm eff} = m/ \gamma^2$ with $\gamma^2 = 1- s^2 / 32$, the dispersion relation reads
\begin{align}
\epsilon_\bk &= \frac{\hbar^2}{2m} \left( k_x^2 + (k_y \cos \eta + k_z \sin\eta)^2 \right. \nonumber \\
& \qquad \left. + \gamma^2 (k_z \cos \eta - k_y \sin\eta)^2 \right)\, . 
\label{eq:dispersion_effective_mass}
\end{align} 
The correction to the ground-state energy can now be calculated according to (\ref{eq:pines}) with the difference that we integrate over the whole momentum space and use (\ref{eq:dispersion_effective_mass}) and (\ref{eq:ddi_ft}) for the single-particle dispersion and interaction potential, respectively. The integrals are most conveniently performed in spherical coordinates and the beyond-mean-field corrections for a dipolar Bose gas in the presence of a weak one-dimensional lattice read as (see Appendix~\ref{app:weak} for details)

\begin{align}
\frac{E_0^{(2)}}{V} &= \frac{8}{15 \pi^2} \frac{(ng)^{5/2} m^{3/2}}{\hbar^3} \Bigg\lbrace F(\varepsilon_{dd}) \nonumber \\
& \qquad  \left. +\frac{s^2}{64} \left[
F(\varepsilon_{dd}) + \frac{1}{2} (3 \cos^2 \eta -1) H(\varepsilon_{dd}) \right] \, \right\}\, 
\label{eq:LHY_weak}
\end{align}
with $F(\varepsilon_{dd}) = \frac{1}{2} \int_{-1}^1 du \,  (1 + \varepsilon_{dd} (3u^2 - 1))^{5/2}$ and $H(\varepsilon_{dd}) = \frac{1}{2} \int_{-1}^1 du \, (1+\varepsilon_{dd} (3u^2 - 1))^{5/2} (3u^2 -1)$. 

\begin{figure}
\centering
\includegraphics[width=0.45\textwidth]{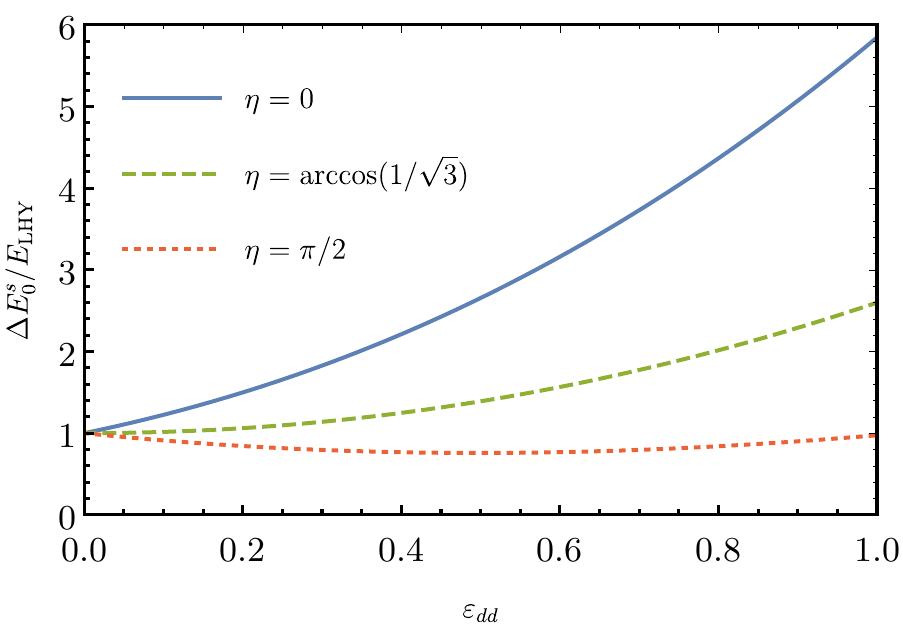}
\caption{Leading-order correction to the beyond-mean-field contribution of the ground-state energy as a function of the relative dipole interaction strength $\varepsilon_{dd}$ for different tilting angles $\eta$ between the polarization axis of the dipoles and the wave vector of the lattice. The angle $\eta = 0$ [blue (solid) line] corresponds to the case where the lattice wave vector is parallel to the dipoles, the angle $\eta = \pi/2$ [red (dotted) line] corresponds to the case where the lattice is applied perpendicular to the dipoles. At $\eta = \arccos(1/\sqrt{3}) \approx 54.7\, {}^\circ$, the anisotropic correction vanishes and only the isotropic correction contributes [green (dashed) line].}
\label{fig:eqf_effmass_arbitrary}
\end{figure}

The energy correction thus has the following form: The first term inside the curly brackets is the usual LHY correction in the case of a dipolar gas in free space~\cite{Lima2012}. The second term accounts for the influence of the lattice and has two parts. The first term is isotropic, has the same structure as in free space and can be explained by an effective mass in one direction. The other part proportional to $H(\varepsilon_{dd})$ is clearly anisotropic in the sense that it depends on the orientation of the lattice with respect to the dipoles. In Fig.~\ref{fig:eqf_effmass_arbitrary}, we plot the function $F(\varepsilon_{dd}) + 1/2 (3 \cos^2 \eta - 1) H(\varepsilon_{dd})$ for different values of the tilting angle $\eta$ between lattice wave vector and polarization of the dipoles. When the lattice is oriented parallel to the dipoles [$\eta = 0$, blue (solid) line], the corrections are enhanced and monotonically increase with increasing $\varepsilon_{dd}$. In this setup, the dipoles are directed by the trapping potential to arrange in a side-by-side configuration so that the fluctuations have mainly repulsive character. On the other hand, when the lattice is orientated perpendicular to the dipoles [$\eta = \pi/2$, red (dotted) line], the correction first decreases for small $\varepsilon_{dd}$, reaches a minimum, and finally increases for larger $\varepsilon_{dd}$. In contrast to the parallel orientation, the dipoles are now dragged to the head-tail configuration such that the attractive character of the fluctuations is enhanced. In this case, the correction is also much smaller than in the case where the lattice is oriented parallel to the dipoles. At the "magic angle" $\eta = \arccos(1/\sqrt{3}) \approx 54.7\, {}^\circ$, the anisotropic correction vanishes and only the isotropic correction contributes [green (dashed) line]. 

\subsection{Mean-field terms}

In the above discussion, we have so far neglected the contributions coming from the mean-field terms which also have an isotropic and anisotropic part where the latter comes from the orientation of the lattice relative to the dipoles. The mean-field terms can be written as
\begin{equation}
\frac{E_\text{MF}}{V} = \frac{1}{2}n^2 g + \frac{s^2}{64} \left( \frac{n^2 g}{2} (1 +   \varepsilon_{dd}(3 \cos^2 \eta - 1))\right) \, . 
\end{equation}
The first term is the usual contribution from the chemical potential in free space in the absence of any lattice. The second part is the leading-order correction to the mean-field energy in the presence of a weak lattice and provides an anisotropic correction. The anisotropic correction has the same functional dependence on the angle between the lattice wave vector and the polarization axis of the dipoles but, apart from the different scaling in the density, a different functional dependence on the relative dipole interaction strength $\varepsilon_{dd}$. While the mean-field term goes linearly with $\varepsilon_{dd}$ for all values, the function $H(\varepsilon_{dd}) \approx 2\varepsilon_{dd} + 6/7 \varepsilon_{dd}^2$ is linear only for small values of $\varepsilon_{dd}$.

\section{Conclusions}
\label{sec:conclusions}

In this paper, we studied the effects of an optical lattice on the beyond-mean-field corrections for a dipolar Bose gas with emphasis on the ability to control and manipulate the strength of these corrections with respect to the depth  and the orientation of the lattice. In the case of a deep three-dimensional lattice, the presence of the lattice introduces a nontrivial density dependence of the beyond-mean-field terms whose form is independent of the strength of the dipolar interaction but whose strength can be enhanced by increasing the dipolar interaction strength. For a weak one-dimensional lattice, we find that the strength of the beyond-mean-field corrections can be controlled by manipulating the relative orientation of the lattice and the dipole axis. In view of the current experiments on dipolar quantum droplets, we would like to point out that our results of the study of the deep optical lattice are not directly applicable to these droplets due to the missing cancellation of the mean-field contributions. However, our results for the weak one-dimensional case might be of importance for future experiments with droplets in weak lattices.

\section{Acknowledgements}
This work was supported by the Deutsche Forschungsgemeinschaft (DFG) within the research unit FOR 2247.

\appendix

\section{Dipolar interactions in a deep lattice}
\subsection{Lattice Fourier transform}
\label{app:appendix_dipolarFT}

In this section, we present the calculation of the discrete Fourier transform of the dipolar potential as given in \eref{eq:ft_potential}. The calculation is based on the analogous one performed in two dimensions~\cite{PeterPhD} extended to three dimensions. Note that in this section, we denote the summation over lattice sites by $\sum_{\mathbf{R}}$ and measure all lengths and momenta in terms of the lattice spacing and the inverse lattice spacing, respectively.

The lattice Fourier transformation of the dipolar interaction can be written as
\begin{equation}
V_{dd}(\mathbf{k}) = g \, \varepsilon_{dd} \frac{3}{4\pi} \sumRp  \frac{R^2-3z^2}{R^5}e^{i \mathbf{k} \mathbf{R}}\, ,
\end{equation}
which we may rewrite as
\begin{equation}
V_{dd}(\mathbf{k}) = g \, \varepsilon_{dd} \frac{3}{4\pi}  \left(\chi^3(\mathbf{k}) + 3 \frac{\partial^2}{\partial k_z^2} \chi^5(\mathbf{k}) \right)\, ,
\end{equation}
with 
\begin{equation}
\chi^n(\mathbf{k}) = \sumRp \frac{1}{R^n} e^{i \mathbf{k} \mathbf{R}}\, .
\end{equation}
Note that in three dimensions, both $\chi^3(\mathbf{k})$ as well as $ \frac{\partial^2}{\partial k_z^2} \chi^5(\mathbf{k})$ are divergent for $\mathbf{k} \to 0$. However, as will be shown, the sum of both contributions leads to a finite result which is non-analytic for $\mathbf{k} \to 0$. 
\subsection*{Calculation of $\chi^n$}
In order to calculate $\chi^n$, we note the following identity
 \begin{equation}
 \frac{1}{R^n} = \frac{1}{\Gamma(n/2)} \int\limits_0^\infty d u \, e^{-uR^2} u^{\frac{n}{2}-1}, \quad n>0\, ,
 \end{equation}
where $\Gamma(m)$ denotes the Gamma function. In the end, we will be interested in $n = 3$ and $5$.
Using the above identity and splitting up the integral into regions from $0$ to $\eta$ and $\eta$ to $\infty$, we arrive at
\begin{align}
\sumRp \frac{1}{R^n} e^{i \mathbf{k} \mathbf{R}} &= \frac{1}{\Gamma(n/2)} \sumRp \left(\int\limits_0^\eta + \int\limits_\eta^\infty \right) d u \, e^{-uR^2} u^{\frac{n}{2}-1} e^{i \mathbf{k} \mathbf{R}} \nonumber \\
&= \frac{\eta^{n/2}}{\Gamma(n/2)} \sumRp \left(\int\limits_1^\infty  du \, e^{-\frac{\eta R^2}{u}} u^{-\frac{n}{2}-1} \right. \nonumber \\
& \qquad \left. + \int\limits_1^\infty du \, e^{-\eta u R^2} u^{\frac{n}{2}-1} \right) e^{i \mathbf{k} \mathbf{R}} \, .
\label{eq:1}
\end{align}
In the second step, we have rescaled $u$ by $\eta/ u$ for the first integral and by $\eta u$ in the second one. The parameter $\eta$ is the so-called Ewald parameter and determines the boundary between the summation in real space and the summation in the momentum space. In the end, this parameter should be chosen such that convergence is achieved rapidly for both sums. However, the result is independent of the choice of $\eta$.

Next, we use Poisson's summation formula to turn the sum of the first part in (\ref{eq:1}) into a sum in momentum space. In $d$ dimensions, Poisson's summation formula applied to our case reads as
\begin{equation}
\sumRp e^{-a R^2} e^{i \mathbf{k} \mathbf{R}} = \sumR e^{-a R^2} e^{i \mathbf{k} \mathbf{R}} - 1 = \frac{\pi^{d/2}}{a^{d/2}} \sumq e^{- \frac{\vert \mathbf{q} + \mathbf{k} \vert^2}{4 a}} - 1\, .
\end{equation}
With $a = \eta / u$ and $d = 3$, we arrive at
\begin{align}
\chi^n(\mathbf{k}) &= \frac{\eta^{n/2}}{\Gamma(n/2)} \left( \frac{\pi^{3/2}}{\eta^{3/2}} \sumq \int\limits_1^\infty du \, u^{- \frac{n}{2} + \frac{1}{2}} e^{- \frac{u}{4\eta} \vert \mathbf{q} + \mathbf{k} \vert^2} - \frac{2}{n}  \right. \nonumber \\
& \quad \left.  + \sumRp \int\limits_1^\infty du \, e^{-\eta u R^2} u^{\frac{n}{2} -1} e^{i \mathbf{k} \mathbf{R}} \right) \nonumber \\
&=\frac{\eta^{n/2}}{\Gamma(n/2)} \left( \frac{\pi^{3/2}}{\eta^{3/2}} \sumq E_{\frac{n-1}{2}}\left(\frac{\vert \mathbf{q} + \mathbf{k} \vert^2}{4\eta} \right) - \frac{2}{n} \right. \nonumber \\
& \quad \left. + \sumRp E_{\frac{2-n}{2}}(\pi R^2) e^{i \mathbf{k} \mathbf{R}} \right) 
\, 
\end{align}
with the exponential integral function
\begin{equation}
E_m(x) = \int\limits_1^\infty \frac{e^{-xt}}{t^m}\, . 
\end{equation}
Setting $n = 3$ and choosing $\eta = \pi$, we get
\begin{align}
\chi^3(\mathbf{k}) & = 2\pi \left( \sumq E_{1}\left(\frac{\vert \mathbf{q} + \mathbf{k} \vert^2}{4\pi} \right) - \frac{2}{3} \right. \nonumber \\
& \qquad \qquad  \left. + \sumRp E_{-1/2}(\pi R^2) e^{i \mathbf{k} \mathbf{R}} \right) \, .
\end{align}
Note that the function $E_1(x)$ diverges logarithmically for $x \to \infty$. This is expected as the sum $\sumRp 1/R^3$ diverges logarithmically in three dimensions. 

\subsection*{Calculation of the anisotropic part}
Now we turn to the anisotropic part given by $\frac{\partial^2}{\partial k_z^2} \chi^5(\mathbf{k})$.
For $\eta = \pi$ and $n  =5$, we get
\begin{align}
\chi^5(\mathbf{k})&= \frac{4\pi^2}{3} \left( \sumq E_{2}\left(\frac{\vert \mathbf{q} + \mathbf{k} \vert^2}{4\pi} \right) - \frac{2}{5} \right. \nonumber \\
&\qquad  \qquad \left. + \sumRp E_{-3/2}(\pi R^2) e^{i \mathbf{k} \mathbf{R}}\right)\, .
\end{align}
Differentiating the last term twice with respect to $k_z$, we obtain
\begin{equation}
\frac{\partial^2}{\partial k_z^2} \sumRp E_{-3/2}(\pi R^2) e^{i \mathbf{k} \mathbf{R}} = - \sumRp z^2\,  E_{-3/2}(\pi R^2) e^{i \mathbf{k} \mathbf{R}}\, .
\end{equation}
For the first term, we note the relation
\begin{equation}
E'_n(x) = - E_{n-1}(x)
\end{equation}
and thus have
\begin{align}
\frac{\partial^2}{\partial k_z^2} \sumq E_{2}\left(\frac{\vert \mathbf{q} + \mathbf{k} \vert^2}{4\pi} \right)  & = \sumq E_0 \left(\frac{\vert \mathbf{q} + \mathbf{k} \vert^2}{4\pi} \right)  \frac{(k_z + q_z)^2}{4\pi ^2}  \nonumber \\
& \qquad - E_1 \left( \frac{\vert \mathbf{q} + \mathbf{k} \vert^2}{4\pi} \right)  \frac{1}{2\pi}\,  .
\end{align}
Finally, we end up with
\begin{align}
\frac{\partial^2}{\partial k_z^2} &\chi^5(\mathbf{k}) = \frac{4\pi^2}{3} \left(\sumq E_0 \left(\frac{\vert \mathbf{q} + \mathbf{k} \vert^2}{4\pi} \right)  \frac{(k_z + q_z)^2}{4\pi ^2} \right. \nonumber \\
&   \left. - E_1 \left( \frac{\vert \mathbf{q} + \mathbf{k} \vert^2}{4\pi} \right)  \frac{1}{2\pi}  - \sumRp z^2\,  E_{-3/2}(\pi R^2) e^{i \mathbf{k} \mathbf{R}} \right)\, .
\end{align}
In this expression, we again encounter the diverging part $E_1(x)$ with exactly the prefactor needed in order to cancel the contribution from the isotropic $1/R^3$ part. 

Combining everything and using the explicit form of $E_0(x) = e^{-x} / x$, the final result reads as
\begin{align}
V_{dd}(\mathbf{k}) &=  g \, \varepsilon_{dd} \left( \left(3 e^{-\frac{k^2}{4\pi}} \frac{k_z^2}{k^2} - 1\right)  + \frac{3}{2}\sumRp  E_{-1/2}(\pi R^2) \right. \nonumber \\
& \left. - 2\pi z^2 E_{-3/2} (\pi R^2) e^{i \mathbf{k} \mathbf{R}}  + 3 \sumqp e^{- \frac{\vert \mathbf{q}  + \mathbf{k} \vert ^2}{4 \pi}} \frac{(k_z + q_z)^2}{\vert \mathbf{k} + \mathbf{q} \vert^2} \right) .
\label{eq:vk_final}
\end{align}
The first part exactly reproduces the continuous Fourier transform of $V_{dd}(\mathbf{k})$ in the limit $\mathbf{k} \to 0$, while the second part, including the sums, vanishes in this limit as is shown in the next paragraph. Note in addition that this expression can be evaluated numerically very efficiently since the summations converge very quickly. For numerical purposes, it is useful to replace $\mathbf{q} = 2 \pi \mathbf{R}$ in the second summation such that convergence is achieved using only a few lattice sites.

\begin{figure}
\centering
\includegraphics[width=0.45\textwidth]{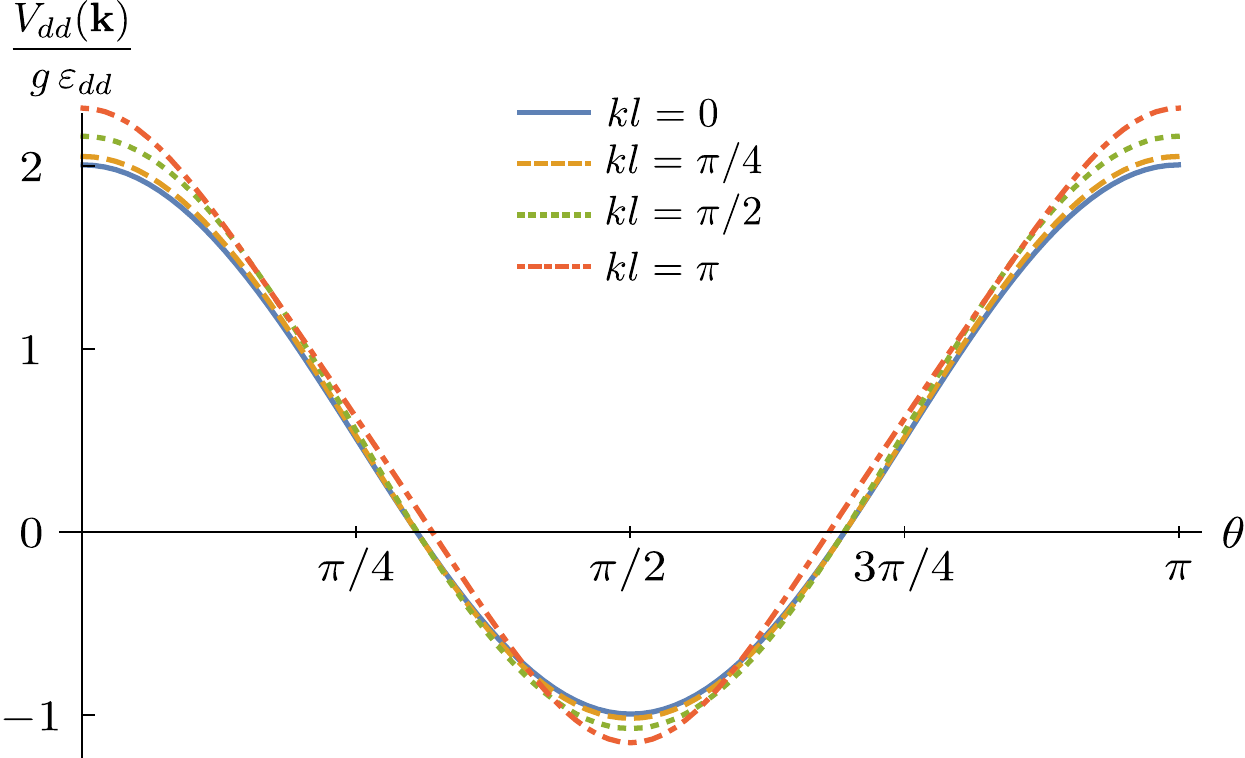}
\caption{Angular dependence of the lattice Fourier transform $V_k$ for $ kl = 0$ [blue (solid) line], which corresponds to the result in free space, $kl = \pi/4$ [orange (dashed) line], $kl = \pi/2$ [green (dotted) line], and $kl = \pi$ [red (dash-dotted) line].}
\label{fig:Vk_diffk}
\end{figure}

\begin{figure}
\includegraphics[width=0.45\textwidth]{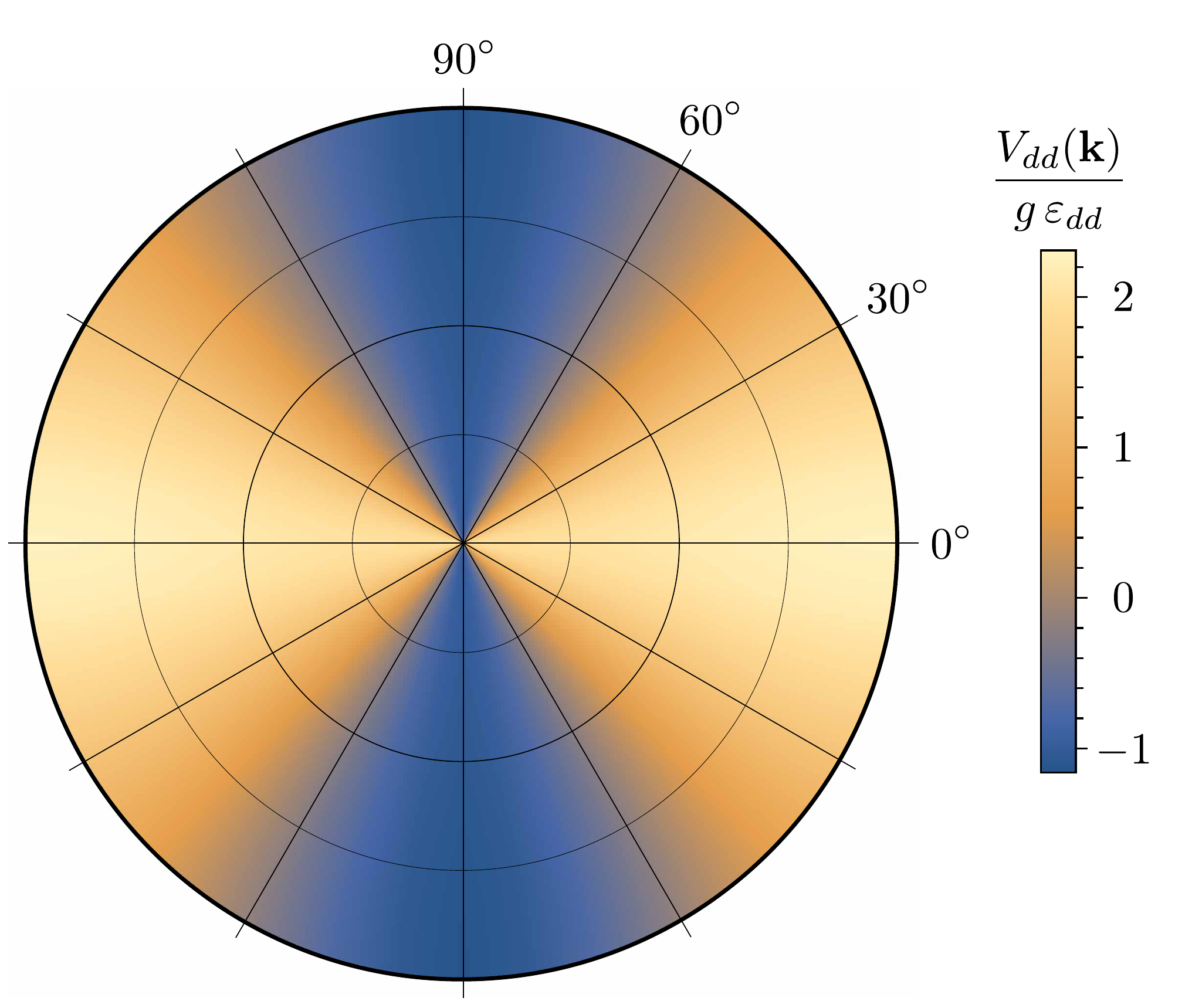}
\caption{Polar plot of $V_{dd}(k, \theta, \phi = 0)$ for $kl \in [0, \pi]$. The boundary circle corresponds to $k l = \pi$ while the inner circles correspond to $ k l = \pi / 4, \, \pi / 2 , \, 3 \pi / 4$ (from innermost circle towards to edge).}
\label{fig:Vk_polar}
\end{figure}

In Figs.~\ref{fig:Vk_diffk} and \ref{fig:Vk_polar}, the angular dependence of \eref{eq:vk_final} is plotted for different values of $k$ and one can see that in contrast to the free space result, the Fourier transform also depends on the absolute value of $\mathbf{k}$. The corrections to the free-space Fourier transform become even more apparent when plotting both functions over the first Brillouin zone of a simple cubic lattice along paths of high symmetry as shown in Fig.~\ref{fig:Vk_BZ}. Around $\mathbf{k} = 0$, there is little difference between both functions, whereas for finite momenta, deviations become clearly visible since the anisotropic structure of the lattice is probed.

\subsection*{Continuous limit $\mathbf{k} \to 0$ and non-analytic behavior}
In the long wavelength limit $\mathbf{k} \to 0$, the first term in (\ref{eq:vk_final}) reduces to the continuous Fourier transform $V_{dd}(\mathbf{k})  = g \, \varepsilon_{dd}(  3 \cos^2 \theta - 1) $, where $\theta$ is the angle between $\mathbf{k}$ and the $z$-axis, which is assumed to be the axis along which the dipoles are polarized. 

The other terms in (\ref{eq:vk_final}) for $\mathbf{k} \to 0 $ read as
\begin{align}
\frac{3}{2}&\sumRp \left( E_{-1/2}(\pi R^2) - 2\pi z^2 E_{-3/2} (\pi R^2) \right)  + 3\sumqp e^{- \frac{q^2}{4 \pi}} \frac{ q_z^2}{q^2} \nonumber \\
 &= \frac{3}{2}\sumRp \left( E_{-1/2}(\pi R^2) - 2\pi z^2 E_{-3/2} (\pi R^2) \right)  \nonumber \\
 & \qquad \qquad + 3 \pi \sumRp z^2 \, E_0(\pi R^2)\, ,
\end{align}

where we replaced $\mathbf{q} = 2 \pi \mathbf{R}$. In order to simplify this expression further, we make use of the recurrence relation
\begin{equation}
E_0(x) = E_m(x) + \frac{m}{x}E_{m+1}(x)\, .
\end{equation}
Choosing $m = 3/2$, we get
\begin{equation}
\frac{3}{2} \sumRp E_{-1/2} (\pi R^2) \left( 1 - 3 \frac{z^2}{R^2} \right)  = 0\, 
\end{equation}
due to the symmetry of the lattice. Thus, we end up with 
\begin{equation}
V_{dd}(\mathbf{k}  \to 0) = g\, \varepsilon_{dd} \left( 3\frac{k_z^2}{k^2} - 1 \right) = g \, \varepsilon_{dd} \left( 3 \cos^2 \theta -1 \right) \, .
\end{equation}
This is exactly the same result as in free space and is non-analytic for $\mathbf{k} \to 0$.


\subsection{Role of the next-order terms}
\label{app:nextorder}
For the analysis of the beyond-mean-field terms in the main text, we only included the density-density interactions. However, dipolar interaction gives rise to terms like density-induced tunneling or pair hopping. Including the interaction-induced nearest-neighbor couplings results in~\cite{Sowinski2012,Wall2013,Dutta2015}
\begin{align}
H &= -t \sum_{\langle i, j \rangle} b^\dagger_i b_j + \frac{U}{2} \sum_i n_i(n_i-1) + \frac{V}{2} \sum_{\langle i, j \rangle } n_i n_j \nonumber \\
& \qquad \qquad - T \sum_{\langle i, j \rangle} b^\dagger_i (n_i + n_j) b_j + \frac{P}{2} \sum_{\langle i,j \rangle} {b_i^\dagger}^2 b_j^2 \, .
\label{eq:extendedBHM}
\end{align}
Transforming this to the quasi momentum space, one obtains
\begin{align}
H = \sum_\bk \epsilon_\bk b_\bk^\dagger b_\bk + \frac{1}{N_L}\sum_{\bk, \bk', \bq, \bq'} \delta_{\bk-\bk'+\bq-\bq',\mathbf{K}_m} &f( \bq, \bq', \bk') \nonumber \\ \times &b^\dagger_\bk b^\dagger_\bq b_{\bk'} b_{\bq'}  \, 
\end{align}
with
\begin{align}
f( \bq, \bq', \bk’)&=\frac{U}{2} + \frac{V}{2} \sum_{\boldsymbol{\delta}} e^{- i \boldsymbol{\delta} (\bq-\bq')} \nonumber \\
&- T \sum_{\boldsymbol{\delta}} \left( e^{i \boldsymbol{\delta} \bq'} +  e^{- i \boldsymbol{\delta} (\bq - \bq' - k')}\right)  + \frac{P}{2} \sum_{\boldsymbol{\delta}} e^{i \boldsymbol{\delta}(\bk' + \bq’)}.
\end{align}
Here, the sums over $\boldsymbol{\delta}$ are perfumed over nearest neighbors only. This Hamiltonian has the same structure as the one without additional terms but with modified effective interaction $f( \bq, \bq', \bk’)$. We can thus perform the Bogoliubov approximation, assuming the presence of the condensate at zero momentum and replacing $b_0\to\sqrt{N_0}$, where $N_0$ is the number of particles in the condensate with $N=N_0+\frac{1}{2}\sum^\prime_\bp{b^\dagger_\bp b_\bp+b^\dagger_{-\bp}b_{-\bp}}$, where the prime denotes omitting the zero-momentum mode. For the case of 3D cubic lattice this gives
\begin{widetext}
\begin{align}
H = \frac{N^2}{N_L} f(0,0,0) + \frac{1}{2}\sideset{}{'}\sum_\bp \left\{ \epsilon_\bp \left(b_\bp^\dagger b_\bp + b_{-\bp}^\dagger b_{-\bp} \right) - 2\frac{N}{N_L} f(0,0,0)\left( b_\bp^\dagger b_\bp + b_{-\bp}^\dagger b_{-\bp}  \right) \right. \nonumber \\
+ \frac{N}{N_L} \left[ f(\bp,\bp,0) + f(\bp,0,\bp) + f(0,\bp,0) + f(0,0,\bp) \right] \left(b_\bp^\dagger b_\bp + b_{-\bp}^\dagger b_{-\bp} \right)  \nonumber \\
\left. + 2\frac{N}{N_L} \left( f(0,-\bp,\bp) b_\bp b_{-\bp} + f(-\bp,0,0) b_\bp^\dagger b_{-\bp}^\dagger \right) \right\} \, .
\end{align}
\end{widetext}
The relevant values for a cubic lattice in 3D are
\begin{align}
f(0,0,0) &= \frac{U}{2} + 3V - 12 T + 3P\, , \\
f(\bp,\bp,0) &= \frac{U}{2} + 3V - T( 2 c_\bp + 6) + P c_\bp\, ,\\
f(\bp,0,\bp) &= \frac{U}{2} + V c_\bp - 12 T + P c_\bp \, ,\\  
f(0,0,\bp) &= \frac{U}{2} + 3 V - T (2 c_\bp + 6) + P c_\bp\, ,\\
f(0,\bp,0) &= \frac{U}{2} + V c_\bp - 4T c_\bp + P c_\bp\, ,\\
f(0,-\bp,\bp) &= \frac{U}{2} + V c_\bp - T ( 2 c_\bp +6) + 3P \, , \\
f(-\bp,0,0) &= \frac{U}{2} + V c_\bp - T( 2c_\bp + 6) + 3P\, .
\end{align}
Here, we defined $\sum_{\boldsymbol{\delta}} e^{i \boldsymbol{\delta} \bp} \equiv c_\bp$ and $c_{-\bp} = c_\bp$ for a cubic lattice.

Now, we perform the Bogoliubov transformation and arrive at the formula for the ground state energy
\begin{equation}
E_0 = E_{\text{MF}} + \sideset{}{'} \sum_\bp (E_\bp - \alpha_\bp) \, .
\end{equation}
with
\begin{equation}
E_\bp \equiv \sqrt{\alpha_\bp^2 - \beta_\bp^2} \, 
\end{equation}
and
\begin{align}
\alpha_\bp = c_\bp(-2t - 8 nT  + 4nP + 2nV) + 6t + nU - 6nP\, ,\\
\beta_\bp = nU + 2 n c_\bp (V-2T) - 12nT + 6 nP\, . 
\end{align}
Defining
\begin{align}
\tilde{U}_\bp = U - 12T + 6P + 2c_\bp(V-2T)\, , \\
\tilde{\epsilon}_\bp = -2(t + 2nT - nP)(c_\bp - 3)\, ,
\end{align}
the Bogoliubov spectrum takes the form
\begin{equation}
E_\bp = \sqrt{\tilde{\epsilon}_\bp ( \tilde{\epsilon}_\bp + n \tilde{U}_\bp)}
\end{equation}
similar to the standard Bose-Hubbard model, where $\tilde{\epsilon}_\bp = - 2 t (c_\bp - 3)$ and $\tilde{U}_\bp = U$. One can also see that the spectrum is gapless and linear for small momenta
\begin{equation}
E_\bp = \sqrt{2 \tilde{t} \tilde{U}_0} \vert \mathbf{p} \vert
\end{equation} 
with the renormalized hopping amplitude $\tilde{t} = t + 2nT - nP$ which is now density dependent. The sound velocity is given by $ c = \sqrt{2 \tilde{t} \tilde{U}_0}$. The renormalized hopping amplitude also renormalizes the effective mass which is now given by $m_{\rm eff} = \hbar^2 / 2 l^2 \tilde{t}$. We see that the modifications resulting from additional terms do not fundamentally change the properties of the superfluid. In our case the lattice is assumed to be deep and the role of tunneling terms in the interaction is perturbative.

\section{Dispersion in the presence of a weak optical lattice}
\label{app:weak}

In this appendix, we first state some important results from the physics of a single particle with mass $m$ in the presence of a weak optical lattice. We restrict our discussion to the case of a one-dimensional lattice along the $z$-axis of the form
\begin{equation}
U_L(z) = U \sin^2(q_L z) = \frac{U}{2}\left[1 + \cos(2 q_L z)\right]\, ,
\end{equation}
where $q_L = \pi / l$ is the lattice vector and $l$ is the lattice period. The energy scale of the lattice is given by the recoil energy $E_R = \hbar^2 q_L^2 / 2m$. The full spectrum of the resulting single-particle Hamiltonian can be obtained by diagonalization which is in general only possible numerically. However, since we are interested only in weak lattices with $s = U/ E_R \ll 1$, we can calculate the spectrum analytically using perturbation theory in the parameter $s$. In second order, the correction to the free-space energy is given by
\begin{equation}
\epsilon^s_k = \epsilon^0_{k-K} + \sum_{K'} \frac{\vert U_{K' - K} \vert^2}{\epsilon^0_{k - K} - \epsilon^0_{k - K'}} + \mathcal{O}(U^3) \, ,
\label{eq:dispersion_PT}
\end{equation}
where $K$ is a vector of the reciprocal lattice, $K = 2 n q_L$ with integer $n$, the free-space dispersion $\epsilon^0_k = \hbar^2 k^2 / 2m$, and the Fourier transform of the lattice potential, $U_K = U/2 (\delta_{K, 2q_L} + \delta_{K, -2q_L})$. Note that we omit the constant shift $U/2$ which leads to a shift in the chemical potential and in the end, we choose the dispersion such that $\epsilon^s_{k = 0} =~0$. Setting $k = z q_L$, $z \in [-1,1]$, we can write
\begin{align}
\epsilon_z^s &= E_R \left((z - 2n)^2 + \frac{s^2}{16} \left( \frac{1}{(z-2n)^2 - (z-2(n+1))^2} \right. \right. \nonumber \\
& \quad \left. \left. + \frac{1}{(z-2n)^2 - (z - 2(n-1))^2}\right) \right) + \mathcal{O}(s^3) \, ,
\label{eq:dispersion_nondeg}
\end{align}
where $n$ denotes the index of the Bloch band.

Note that (\ref{eq:dispersion_PT}) only holds for non-degenerate energies away from the edges and the center of the Brillouin zone and in a region of energies where $\vert\epsilon^0_{k - K} - \epsilon^0_{k - K'}\vert \gg U_K$. Close to the edges of the Brillouin zone, $k = \pm q_L$, energies are degenerate and non-degenerate perturbation theory cannot be applied. The dispersion relation in this case reads as
\begin{align}
\epsilon_k^s &= \frac{\epsilon^0_k - \epsilon^0_{k-K}}{2} \pm \sqrt{\left( \frac{\epsilon^0_k - \epsilon^0_{k - K}}{2} \right)^2 + \frac{\vert U_K \vert^2}{4}} \\
&= E_R \left(\frac{z^2 - (z-2n)^2}{2} \pm \sqrt{ \left(\frac{z^2 - (z-2n)^2}{2} \right)^2 + \frac{s^2}{4}} \right)\, . 
\label{eq:dispersion_degenerate}
\end{align}

The above results suggest a splitting into a degenerate and non-degenerate region for the lowest two bands where the border of both regions is determined by the condition $\epsilon^0_{k - K} - \epsilon^0_{k - K'} = U_K$ leading to $z = 1 - s/8$. In the following, we will assume that $z$ is positive but not restricted to values smaller than $1$ as indicated above. Thus, the integration has to be split into three regions. The two non-degenerate regions are from $z = 0$ to $1-s/8$ and from $z = 1+s/8$ to $\infty$ \footnote{Note that to leading order we only consider a splitting up between the lowest and the second lowest Bloch band at $k = \pm q_L$ ($z = \pm 1$) while all other splittings are ignored.}. The degenerate region extends from $z = 1 - s/8$ to $1+ s/8$. 

\subsection{Contribution from the degenerate region}
Here, we argue that the contribution from the degenerate region will not contribute in leading order in $s$. In order to see that, we expand (\ref{eq:dispersion_degenerate}) for $z^2 - (z - 2n)^2 \ll s^2$ which leads to
\begin{align}
\epsilon_k^s = \frac{E_R}{2} \left(  [z^2 + (z - 2)^2] - \frac{s}{8} - \frac{16(z - 1)^2}{s}  \right)
\end{align}
for $1-s/8 \leq z \leq 1$, and
\begin{align}
\epsilon_k^s = \frac{E_R}{2} \left( [(z^2 + (z - 2)^2] + \frac{s}{8} + \frac{16(z - 1)^2}{s}  \right)
\end{align}
for $1 \leq z \leq 1+s/8$.

One can now show that the contribution to the integral (\ref{eq:integral_contact}) coming from the degenerate region is of order $s^3$. Next, we show that the leading-order corrections are indeed of order $s^2$ and come exclusively from the non-degenerate region.

\subsection{Contributions from the non-degenerate region}
For the non-degenerate region which ranges from $z = 0$ to $1-s/8$ and from $z = 1+s/8$ to $\infty$, we have to consider an integral of the form 
\begin{equation}
I(s) = \int_0^{1-s/8} dz \, h(z,s) + \int_{1+s/8}^\infty dz \, h(z,s)\, ,
\end{equation}
where $h(z,0) = 0$. For small $s$, we can expand the integral 
\begin{equation}
I(s) \approx I(0) + s \,\partial_s I(s) \vert_{s = 0} + \frac{1}{2}s^2\,
  \partial_s^2 I(s) \vert_{s = 0}\, .
\end{equation}
The constant term $I(0)$ as well as the linear term vanish due to the structure of $h(z,s)$. The quadratic term, using again that $h(z,0) = 0$, yields
\begin{equation}
\partial_s^2 I(s) \vert_{s = 0} = \text{P.V.} \int_{0}^\infty dz\, \partial_s^2 h(z,s) \vert_{s = 0}\, ,
\end{equation}
where P.V. denotes the Cauchy principal value. When evaluating the integral, divergences will arise in the vicinity of $z = 1$ since the non-degenerate approach is no longer valid. However, the divergences for $z >1$ and $z<1$ will cancel each other.

\subsection{Expansion for $z \ll \beta$}

Since we are also interested in the limit $\beta \to \infty$ and claim that this corresponds to essentially having an effective mass in the direction of the lattice, we look for an expansion of Eq.~(\ref{eq:integral_contact}) for $\beta \to \infty$ or, to be more rigorous, for $z \ll \beta$.  The function $f$ in the integrand plays the role of our lattice dispersion. For small arguments, we can expand Eq.~(\ref{eq:dispersion_nondeg}) around $z = 0$ for the lowest band, $n = 0$:
\begin{align}
\epsilon_k^s &= E_R \left( z ^2 + \frac{s^2}{16} \left( \frac{1}{z^2 - (z - 2)^2} + \frac{1}{z^2 - (z + 2)^2} \right) \right) \nonumber \\
&\approx E_R \left(-\frac{s^2}{32} + \left( 1 - \frac{s^2}{32}\right) z^2 \right)\, .
\end{align}
The first term is just the constant shift that will be subtracted by an additional chemical potential.  The second term gives the leading behavior and corresponds to having an effective mass in the $z$ direction, $m_z~=m/(1 - s^2/32)$.

\subsection{Beyond-mean-field corrections in a weak lattice}
\subsubsection{Contact interaction}

In a next step, we present the details on how to calculate the beyond-mean-field corrections for a contact-interacting gas in a weak lattice. For this purpose, we consider (\ref{eq:corrections_weak_integral}) and, in what follows, we set $\tilde{\epsilon}_k = E_R f(kl)$, where $f(kl)$ is a dimensionless function of the quasi-momentum along the direction of the lattice. Using this form of the lattice dispersion and changing to dimensionless variables $k_\perp \xi = u$ and $kl = z$, we can transform (\ref{eq:corrections_weak_integral}) into
\begin{widetext}
\begin{align}
\frac{\Delta E_0^s}{V} = \frac{1}{2 (2\pi)^3} \frac{(2 m)^{3/2} (n g)^{5/2}}{\hbar^3} 4 \pi \int_0^\infty d u \, u\int_0^\infty d z\,  \sqrt{(u^2 + \beta^2 f(z/\beta))(u^2 +\beta^2 f(z/ \beta) + 2)} &- \sqrt{(u^2 + z^2)(u^2 + z^2 + 2)} \nonumber \\&\quad  - \left(\beta^2 f(z/\beta) -z^2  \right) \, . 
 \label{eq:integral_contact}
\end{align}
\end{widetext}
The factor in front of the integral is proportional to the usual LHY term $\sim m^{3/2} (ng)^{5/2} / \hbar^3$, while the remaining part accounts for the influence of the lattice. As we are now interested in a weak lattice, we calculate the integrals in leading order in $s$. It turns out that the leading order is proportional to $s^2$ and the corrections read
\begin{widetext}
\begin{align}
\frac{\Delta E_0^s}{V} = \frac{1}{2 (2\pi)^3} \frac{(2 m)^{3/2} (n g)^{5/2}}{\hbar^3} 4 \pi \frac{s^2}{2} \frac{\beta^2}{192 \sqrt{2 + \beta^2}} & \left( 8 + 20 \beta^2 + 8 \beta^4 + (4 \sqrt{2}+6 \sqrt{2} \beta^2) \sqrt{2 + \beta^2}  - (4+ 8 \beta^2) (2 + \beta^2) \right.\nonumber \\
& \qquad \left. +3 \beta^2 (2+ \beta^2)( \ln[ \beta( - \beta + \sqrt{2 + \beta^2})] + \ln[\beta(\beta + \sqrt{2 + \beta^2}) ] \right).
\label{eq:corr_contact}
\end{align}
\end{widetext}

\subsubsection{Dipolar interaction}

In this part, we present the details of the calculation of (\ref{eq:LHY_weak}). First, we express the dispersion relation for an anisotropic effective mass in spherical coordinates, that is

\begin{align}
\epsilon_k &= \frac{\hbar^2 k^2}{2m} ( \cos^2 \theta \sin^2 \eta + \sin^2 \theta ( \cos^2 \phi + \gamma^2 \sin^2 \eta \sin^2 \phi) \nonumber \\
& \quad + \cos^2 \eta( \gamma^2 \cos^2 \theta + \sin^2 \phi \sin^2 \theta) \nonumber \\ 
& \quad + (\gamma^2 - 1) \cos \eta \sin \eta \sin \phi \sin 2 \theta ) \nonumber \\
&= \frac{\hbar^2 k^2}{2m} f(\theta, \phi, \gamma, \eta) \, . 
\end{align}
The correction to the ground state energy can now be calculated using (\ref{eq:pines}) according to which
\begin{align}
\frac{E_0^{(2)}}{V} - \frac{1}{2}n &\mu^{(2)} = \frac{1}{2(2\pi)^3} \int_0^\infty dk \, k^2 \nonumber \\
&\times \int d \Omega \,  \frac{\left(\frac{\hbar^2 k^2}{2m} f + nV_k - E_k \right) \left( \frac{\hbar^2 k^2}{2m} f - E_k \right)}{2 E_k} \, , 
\label{eq:hp_ddi}
\end{align}
where $E_k = \sqrt{\epsilon_k (\epsilon_k + 2 n V_k)}$  is the Bogoliubov spectrum and $\int d \Omega = \int_0^\pi d \theta \, \sin \theta \int_0^{2\pi} d \phi$ denotes the integration over the solid angle .

The integral on the right-hand side can be simplified and made dimensionless by pulling out the factor $ng$ and using the healing length $\xi^2 = \hbar^2 / 2 mng$. Making the substitution $(k\xi)^2 f = x^2$ and performing the integral over $k$ leads to
\begin{equation}
\frac{1}{16 \pi^3}\frac{ng}{\xi^3} \left( - \frac{2\sqrt{2}}{15} \right)\int d \Omega\,  \frac{\tilde{V}^{5/2}}{f^{3/2}} \, . 
\end{equation}
The remaining integrals cannot be solved analytically in general. However, we are only interested in the lowest-order corrections which are due to the lattice such that we can expand the function under the integral to the lowest order in $s$, which leads to
\begin{align}
\frac{1}{16 \pi^3}\frac{ng}{\xi^3} &\left( - \frac{2\sqrt{2}}{15} \right)\int d \Omega\,  \tilde{V}^{5/2} \bigg( 1 \nonumber \\
& \left. + \frac{3s^2}{64} ( \cos \eta \cos \theta - \sin \eta \sin \phi \sin \theta)^2\right) \, . 
\end{align}

The first term gives rise to the standard LHY correction induced by dipolar interactions~\cite{Lima2012}, while the latter part includes the corrections due to the lattice. After performing the integral over the angle $\phi$, the second-order correction to the ground state energy in the presence of an optical lattice reads as
\begin{align}
\frac{E_0^{(2)}}{V} &= \frac{8}{15 \pi^2} \frac{(ng)^{5/2} m^{3/2}}{\hbar^3} \bigg( F(\varepsilon_{dd})  \nonumber \\ 
& \qquad  + \frac{3s^2}{512} \int_{-1}^1 du \, (1 + \varepsilon_{dd} (3u^2 - 1))^{5/2} \nonumber \\
& \qquad \times \left. ( u^2 + 1 + (3u^2 -1)\cos 2\eta ) \right)\, 
\end{align}
with $F(\varepsilon_{dd}) = \frac{1}{2} \int_{-1}^1 du \,  (1 + \varepsilon_{dd} (3u^2 - 1))^{5/2}$. The latter term, accounting for the influence of the lattice, can be rewritten as
\begin{align}
\frac{s^2}{64} \left[
F(\varepsilon_{dd}) + \frac{1}{2} (3 \cos^2 \eta -1) H(\varepsilon_{dd}) \right] \, 
\end{align}
with $H(\varepsilon_{dd}) = \frac{1}{2} \int_{-1}^1 du \, (1+\varepsilon_{dd} (3u^2 - 1))^{5/2} (3u^2 -1)$.

\bibliography{main_text_incl_figure_incl_supplement.bbl}

\end{document}